\documentclass[prd,nofootinbib,superscriptaddress,showpacs]{revtex4}
\pdfoutput=1
\usepackage{graphicx}
\usepackage{epsfig}
\usepackage{latexsym}
\usepackage{amsmath}
\usepackage{amsfonts}
\usepackage{amsxtra}
\usepackage{axodraw}

\newcommand{\msbar}{\overline{\mbox{{\sc ms}}}}

\def\krto{ {\,\,\lower .8ex\hbox {$\longrightarrow \atop k \rightarrow 0$}\,\,}}

\def\bea{\begin{eqnarray} }
\def\beq{\begin{eqnarray} }

\def\eea{\end{eqnarray}}
\def\eeq{\end{eqnarray}}
\def\altura#1{\rule[0cm]{0cm}{#1cm}}
\def\eq#1{Eq.~(\ref{#1})}

\begin{document} 

\title{ Ghost-gluon coupling, power corrections and  $\Lambda_{\msbar}$ from lattice QCD \\ 
 with a dynamical charm}

\author{B.~Blossier}
\author{Ph.~Boucaud} 
\affiliation{Laboratoire de Physique Th\'eorique, 
Universit\'e de Paris XI; B\^atiment 210, 91405 Orsay Cedex; France}
\author{M.~Brinet}
\affiliation{Laboratoire de Physique Subatomique et de Cosmologie, CNRS/IN2P3/UJF; 
53, avenue des Martyrs, 38026 Grenoble, France}
\author{F.~De Soto}
\affiliation{Dpto. Sistemas F\'isicos, Qu\'imicos y Naturales, 
Univ. Pablo de Olavide, 41013 Sevilla, Spain}
\author{X.~Du}
\affiliation{Laboratoire de Physique Subatomique et de Cosmologie, CNRS/IN2P3/UJF; 
53, avenue des Martyrs, 38026 Grenoble, France}
\author{M.~Gravina}
\affiliation{Department of Physics, University of Cyprus, P.O. Box 20537, 1678 Nicosia, Cyprus and Computation-based Science and Technology Research Center}
\author{V.~Morenas}
\affiliation{Laboratoire de Physique Corpusculaire, Universit\'e Blaise Pascal, CNRS/IN2P3 
63177 Aubi\`ere Cedex, France}
\author{O.~P\`ene}
\affiliation{Laboratoire de Physique Th\'eorique, 
Universit\'e de Paris XI; B\^atiment 211, 91405 Orsay Cedex; France}
\author{K.~Petrov}
\affiliation{Laboratoire de Physique Th\'eorique, 
Universit\'e de Paris XI; B\^atiment 211, 91405 Orsay Cedex; France}
\author{J.~Rodr\'{\i}guez-Quintero}
\affiliation{Dpto. F\'isica Aplicada, Fac. Ciencias Experimentales; 
Universidad de Huelva, 21071 Huelva; Spain.}

\begin{abstract}

This paper is a first report on the determination of $\Lambda_{\msbar}$ from lattice simulations with 
2+1+1 twisted-mass dynamical flavours {\it via} the computation of the ghost-gluon coupling renormalized 
in the MOM Taylor scheme. We show this approach allows a very good control of the lattice artefacts and 
confirm the picture from previous works with quenched and ${\rm N}_f$=2 
twisted-mass field configurations which prove the necessity to include non-perturbative power corrections 
in the description of the running. We provide with an estimate of $\Lambda_{\msbar}$ in very good agreement with 
experimental results. To our knowledge it is the first calculation with a dynamical charm quark
which makes the running up to $\alpha_s(M_Z)$ much safer. 

\end{abstract}


\maketitle

\begin{flushright}
LPT-Orsay 11-74\\
UHU-FT/11-29 \\
IRFU-11-136
\end{flushright}
\vspace*{-1cm}
\begin{figure}[h]
    \includegraphics[width=25mm]{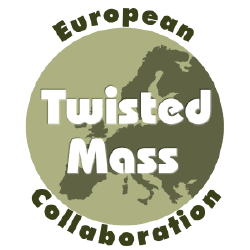}
\end{figure}




\section{Introduction}

QCD,  the theory for the  strong interactions, can be  confronted with 
experiments only after providing it with a few inputs: 
one mass parameter for each quark species and the only surviving parameter 
in the limit of massless quarks, namely $\Lambda_{\rm QCD}$, the energy scale used 
as the typical boundary condition for the integration of the Renormalization Group equation 
for the strong coupling constant. 
Thus, contrary to its running that can be computed in perturbation theory, 
the value of the renormalized strong coupling at any scale, or equivalently $\Lambda_{\rm QCD}$,  
has to be taken from experiment. 

The QCD running coupling can be also obtained from lattice computations, 
where the lattice spacing replaces $\Lambda_{\rm QCD}$ as a 
free parameter to be adjusted from experimental numbers: masses, decay
constants, etc. Different methods have been used for 
the lattice calculation of $\Lambda_{\rm QCD}$. Among the most extensively applied, 
we can enumerate the implemention of the Schr\"odiger functional scheme (see, for
instance,~\cite{Luscher:1993gh,deDivitiis:1994yp,DellaMorte:2004bc,Aoki:2009tf}
and  references therein), those based on the perturbative analysis of
short-distance sensitive lattice observables as the inter-quark static potential 
(see for instance~\cite{Booth:1992bm,Brambilla:2010pp}), heavy-quark potential, Wilson loops or 
small Creutz ratios expanded in the ``boosted'' lattice coupling 
(see~\cite{Gockeler:2005rv,Mason:2005zx,Maltman:2008bx,Davies:2008sw} and
reference therein) or the vacuum polarization functions~\cite{Shintani:2009zz,Shintani:2010ph}; 
and, in particular, those based on the study of  the momentum behaviour of Green functions 
(see~\cite{Alles:1996ka,Boucaud:1998bq,Boucaud:2000ey,Boucaud:2000nd,Boucaud:2001st,Boucaud:2001qz,Sternbeck:2007br}
and references therein).
 In  previous studies we compared the behaviour of the
2-gluon and 3-gluon Green functions as a function of the renormalization scale
with the perturbative predictions. This allowed us
to get  estimations for $\alpha_S$ and 
$\Lambda_{\rm QCD}$; but it also revealed the presence of non-perturbative power corrections
which we  interpreted as coming from the dimension-two
non-zero Landau-gauge gluon condensate in an OPE approach~\footnote{The 
possible phenomenological
implications in the gauge-invariant world of such  a dimension-two Landau-gauge gluon
condensate and in connection with confinement scenarios has been also largely 
investigated, as can be seen for instance in ref.~\cite{Gubarev:2000nz,Dudal:2005na}. This condensate has
been also related to the QCD vacuum properties through the instantons 
liquid model~\cite{Boucaud:2002nc}}. 

In the last few years, several authors of this paper have been pursuing a program to 
study the running of the strong coupling, and so evaluate $\Lambda_{\rm QCD}$, grounded 
on the lattice determination of the ghost-gluon coupling in the so-called MOM Taylor 
renormalization scheme. The main advantage of this ghost-gluon approach is that the 
lattice computation of the coupling only involves the calculation of two-point correlators, which 
yields a very good control of the lattice artefacts over a large momentum window, 
mainly owing to the $H(4)$-extrapolation prescription~\cite{Becirevic:1999uc,deSoto:2007ht},  
and then for a precise checking of the running.
We have first analyzed the pure Yang-Mills case (${\rm N}_f=0$)~\cite{Boucaud:2008gn,RodriguezQuintero:2009jw} 
and next extended the analysis to the case in which
twisted ${\rm N}_f=2$ dynamical quarks were included in the lattice
simulations~\cite{Blossier:2010ky,Blossier:2010we,Pene:2011kg}. 
Now, for the first time, we apply the same approach (outlined in sec.~\ref{sec:proc}) to study the 
strong coupling by dealing with lattice simulations with two light 
degenerate twisted-mass flavours and a heavy doublet 
to include the strange and charm dynamical quarks.
This is done within the framework of the European Twisted Mass (ETM) collaboration from where we 
used several ensembles of gauge fields for different bare lattice couplings,  
 twisted masses and volume to conclude that: 
(i) the running description of the data requires to take into account 
non-perturbative power corrections, which appear to behave as OPE~\cite{Shifman:1978bx,Shifman:1978by} 
predict when a non-vanishing landau-gauge dimension-two gluon condensate 
is considered; 
and (ii) only after taking into account the non-perturbative power corrections, the lattice estimate 
for $\Lambda_{\rm QCD}$ strikingly agrees with the experimental result, as can 
be seen in sec.~\ref{sec:res}.


\section{About the procedure}
\label{sec:proc}

We shall follow the procedure described in detail in
refs.~\cite{Boucaud:2008gn,Blossier:2010ky} to extract an estimate of
$\Lambda_{\msbar}$ from the non-perturbative lattice determination of  $\alpha_T(q^2)$, the
running strong coupling renormalized in the MOM Taylor scheme and Landau gauge.
Let us recall  briefly how the procedure works. The Taylor coupling is defined 
by 
\beq\label{alpha} 
\alpha_T(\mu^2) \equiv \frac{g^2_T(\mu^2)}{4 \pi}=  \ \lim_{\Lambda \to \infty} 
\frac{g_0^2(\Lambda^2)}{4 \pi} G(\mu^2,\Lambda^2) F^{2}(\mu^2,\Lambda^2) \ ,
\eeq
where $F$ and $G$ stand for the bare ghost and gluon dressing functions in Landau-gauge. 
As was thoroughly explained in ref.~\cite{Boucaud:2008gn} 
(see also the appendix A of ref.~\cite{Boucaud:2011ug}), the well-known Taylor's 
paper~\cite{Taylor} proved that, at any order in perturbation, 
the proper ghost-gluon vertex trivially takes its tree-level 
form when the incoming ghost momentum vanishes, this implying that the renormalisation constant 
for this proper ghost-gluon vertex is just equal to 1 in the MOM-like scheme defined by 
the particular kinematics with a zero-momentum incoming ghost, {\it i.e.}  Taylor scheme. 
This is not only true in perturbation but it can be also concluded 
that taking the limit of a vanishing incoming ghost momentum drops 
any non-perturbative correction away from the whole proper ghost-gluon 
vertex, as was discussed in ref.~\cite{Boucaud:2011eh}.
Thus, \eq{alpha} for the Landau-gauge Taylor-scheme running coupling 
can be straighforwardly derived from this last result. 

The ghost and gluon dressing functions will be here  obtained from ${\rm N}_f$=2+1+1 gauge configurations 
for several bare couplings, light twisted masses and volumes. 
Contrarily to the analysis performed in ref.~\cite{Blossier:2010ky}, 
the interplay of light and heavy quark mass and UV cut-off effects  makes a chiral extrapolation harder.
Further studies are underway for a better control of this point.
Thus, for the scope of this paper, we will content ourselves with an estimation of the uncertainty 
due to the quark mass effects.  

It should be emphasized that a crucial
role is played by the appropriate elimination of discretization artefacts to
provide us with reliable and  exploitable results. A first step consists in
curing the artefacts which are due to the breaking of  the rotational invariance on the
lattice, where the remaining symmetry is restricted to the $H(4)$ isometry
group. For this purpose, we perform the so-called $H(4)$-extrapolation 
procedure~\cite{Becirevic:1999uc,Becirevic:1999hj,deSoto:2007ht} that leaves us
with
\beq\label{eq:H4}
\alpha_T^{\rm Latt}\left(a^2p^2,a^2 \frac{p^{[4]}}{p^2},\dots \right) 
\ = \ \widehat{\alpha}_T(a^2 p^2) \ + \ 
\left. \frac{\partial \alpha_T^{\rm Latt}}{\partial \left(a^2
\frac{p^{[4]}}{p^2}\right)}  
\right|_{a^2 \frac{p^{[4]}}{p^2}=0} \ 
a^2 \frac{p^{[4]}}{p^2} \ + \ \dots
\eeq
where $p^{[4]}=\sum_i p_i^4$ is the first $H(4)$-invariant (and the only one indeed
relevant 
in our analysis). Thus, we first average over any combination of momenta being
invariant under 
$H(4)$ ($H(4)$ orbit) and extrapolate then to the "continuum case", where the effect 
of $a^2 p^{[4]}$ must vanish, by applying \eq{eq:H4} for all the orbits sharing the 
same value of $p^2$, with the only assumption that the slope depends smoothly 
on $a^2 p^2$ and can be fitted to a polynomial form from the whole set of lattice
data. 
Furthermore, the $H(4)$-artefact-free lattice coupling, $\widehat{\alpha}_T(a^2 p^2)$ 
might differ from the continuum coupling by some $O(4)$-invariant artefacts, as
shown for example in the lattice analysis of the quark-propagator renormalization
constant~\cite{Boucaud:2003dx,Boucaud:2005rm,Blossier:2010vt}.
This leads us finally to write:
\beq\label{eq:a2p2}
\widehat{\alpha}_T(a^2 p^2) \ = \ \alpha_T(p^2) \ + \ c_{a2p2} \ a^2 p^2 \ + \ {\cal
O}(a^4) \ ,
\eeq
where $c_{a2p2}$ should be fitted from the lattice data and verify the appropriate
scaling 
from the simulations with different bare couplings, $\beta$; while $\alpha_T$ is the
lattice 
prediction to be compared with the continuum OPE formula for the Taylor 
strong coupling~\cite{Blossier:2010ky}, 
\beq\label{alphahNP}
\alpha_T(\mu^2)
\ = \
\alpha^{\rm pert}_T(\mu^2)
\ 
\left( 
 1 + \frac{9}{\mu^2} \
R\left(\alpha^{\rm pert}_T(\mu^2),\alpha^{\rm pert}_T(q_0^2) \right) 
\left( \frac{\alpha^{\rm pert}_T(\mu^2)}{\alpha^{\rm pert}_T(q_0^2)}
\right)^{1-\gamma_0^{A^2}/\beta_0} 
\frac{g^2_T(q_0^2) \langle A^2 \rangle_{R,q_0^2}} {4 (N_C^2-1)}
\right) \ , \nonumber \\
\eeq
where $\gamma_0^{A^2}$ can be taken from \cite{Gracey:2002yt,Chetyrkin:2009kh} to
give for ${\rm N}_f=4$, 
\beq
1-\gamma_0^{A^2}/\beta_0 = \frac {27}{132 - 8 N_f} = \frac{27}{100} \ ;
\eeq
and, applying the same method outlined in the appendix of ref.~\cite{Blossier:2010ky}, 
one can take advantage of the ${\cal O}(\alpha^4)$-computations for the Wilson coefficients in ref.~\cite{Chetyrkin:2009kh}, 
and obtains

\beq
R\left(\alpha,\alpha_0\right) = 
\left( 1 + 1.18692 \alpha + 1.45026 \alpha^2 + 2.44980 \alpha^3 \right) 
\ 
\left( 1 - 0.54994 \alpha_0 - 0.13349 \alpha_0^2 - 0.10955\alpha_0^3 \right) \ ,
\eeq

\noindent
for $q_0=10$ GeV. 
The purely perturbative running in \eq{alphahNP} 
is given up to four-loops by~\cite{Nakamura:2010zzi}
\begin{align}
  \label{betainvert}
  \begin{split}
      \alpha_T^{\rm pert}(\mu^2) &= \frac{4 \pi}{\beta_{0}t}
      \left(1 - \frac{\beta_{1}}{\beta_{0}^{2}}\frac{\log(t)}{t}
     + \frac{\beta_{1}^{2}}{\beta_{0}^{4}}
       \frac{1}{t^{2}}\left(\left(\log(t)-\frac{1}{2}\right)^{2}
     +
\frac{\widetilde{\beta}_{2}\beta_{0}}{\beta_{1}^{2}}-\frac{5}{4}\right)\right)
\\
     &+ \frac{1}{(\beta_{0}t)^{4}}
 \left(\frac{\widetilde{\beta}_{3}}{2\beta_{0}}+
   \frac{1}{2}\left(\frac{\beta_{1}}{\beta_{0}}\right)^{3}
   \left(-2\log^{3}(t)+5\log^{2}(t)+
\left(4-6\frac{\widetilde{\beta}_{2}\beta_{0}}{\beta_{1}^{2}}\right)\log(t)-1\right)\right)
   \end{split}
\end{align}
with $t=\ln \frac{\mu^2}{\Lambda_T^2}$ and the coefficients of the $\beta$-function
in Taylor-scheme~\cite{Chetyrkin:2000dq,Boucaud:2008gn}. 
 As for  the $\Lambda_{\rm
 QCD}$ parameters in Taylor-scheme and $\overline{\rm MS}$, they are  related
through~\cite{Blossier:2010ky}
 \beq\label{ratTMS}
 \frac{\Lambda_{\overline{\rm MS}}}{\Lambda_T} \ = 
 e^{\displaystyle - \frac{507-40 N_f}{792 - 48 N_f}} \ = \ 0.560832 
 \ ,
 \eeq
for the ${\rm N}_f=4$ case~\footnote{
It should be noted that, although $\Lambda_T$ is gauge-dependent, 
$\Lambda_{\overline{\rm MS}}$ is not. Of course, the conversion factor of both parameters to each 
other is also gauge-dependent. Then, as far as this conversion factor can be exactly determined because 
of the RGE invariance of $\Lambda_{QCD}$, the choice of any gauge for the lattice determination 
of $\Lambda_T$ (Landau gauge in our case) is irrespective for the final determination of 
$\Lambda_{\overline{\rm MS}}$.
}. 
Thus, only three parameters, $g^2 \langle A^2 \rangle$,  
$\Lambda_{\overline{\rm MS}}$ and the coefficient for the $O(4)$-invariant
artefacts $c_{a2p2}$, 
remain free to be fitted through the comparison of the prediction given by 
Eqs.~(\ref{eq:a2p2},\ref{alphahNP}) and the lattice estimate of Taylor coupling after 
$H(4)$-extrapolation.


\section{The lattice set-up}
\label{sec:lat}

As already mentioned, we obtain $\alpha_T^{\rm Latt}$ by 
\eq{alpha} from the ghost and gluon propagators computed from the gauge configurations 
simulated at several lattices with ${\rm N}_f$=2+1+1 mass-twisted lattice flavors~\cite{Frezzotti:2000nk} by 
the ETM collaboration~\cite{Baron:2010bv,Baron:2011sf}. In the gauge sector, we use 
the Iwasaki action and compute the propagators as described in
refs.~\cite{Blossier:2010ky}, 
while for the fermion action we have
\beq\label{eq:tmSl}
S_l \ = \ a^4 \sum_x \overline{\chi}_l(x) \left( \altura{0.5} D_W[U] + m_{0,l} + i
\mu_l \gamma_5 \tau_3 \right) \chi_l(x)
\eeq
for the doublet of degenerate light quarks~\cite{Frezzotti:2003xj} and
\beq\label{eq:tmSh}
S_h \ = \ a^4 \sum_x \overline{\chi}_h(x) \left( \altura{0.5} D_W[U] + m_{0,h} 
+ i \mu_\sigma \gamma_5 \tau_1 + \mu_\delta \tau_3 \right) \chi_h(x)
\eeq
for the heavy doublet. $D_W[U]$ is the standard massless Wilson Dirac operator. 
The lattice parameters for the  ensembles of gauge
configurations we 
used are given in tab.~\ref{tab:set-up}. 
Tuning to maximal twist is achieved by choosing a parity odd operator
and determine $\kappa_{crit}$ such that this operator has a vanishing expectation 
value. One appropriate quantity is the PCAC light quark mass and we demand
$m_{PCAC}$ = 0.
We refer the interested reader to
refs.~\cite{Baron:2010bv,Baron:2011sf} 
for more details about the set-up of the twisted mass lattice simulations.

\begin{table}[ht]
\begin{tabular}{|c|c|c|c|c|c|c|}
\hline
$\beta$ & $\kappa_{\rm crit}$ & $a \mu_l$ & $a \mu_\sigma$ & $a \mu_\delta$ &
$(L/a)^3\times T/a$ & confs. \\
\hline
1.95 & 0.1612400 & 0.0035 & 0.135 & 0.170  & $32^3\times 64$ & 50 \\
     & 0.1612400 & 0.0035 &           &           & $48^3\times 96$ & 40 \\
     & 0.1612360 & 0.0055 &           &            & $32^3\times 64$ & 50 \\   
\hline
2.1 &  0.1563570 & 0.0020 & 0.120 & 0.1385 & $48^3\times 96$ & 40 \\  
\hline
\end{tabular}
\caption{Lattice set-up parameters for the ensembles we used in this paper. They correspond with
the ones coded as B35.48, B35.32, B55.32 
and D20.48 in Tab.~1 of ref.~\cite{Baron:2011sf}. The last column stands for the number 
of  gauge field configurations we used.}
\label{tab:set-up}
\end{table}

\section{The results of the analysis}
\label{sec:res}

\subsection{Curing the $H(4)$-artefacts}

\begin{figure}[ht]
  \begin{center}
  \begin{tabular}{cc}
    $F(p^2)$ \rule[0cm]{7.5cm}{0cm} & $\hat{\alpha}_T(p^2)$ \rule[0cm]{7.5cm}{0cm}
    \\
    \includegraphics[width=9cm]{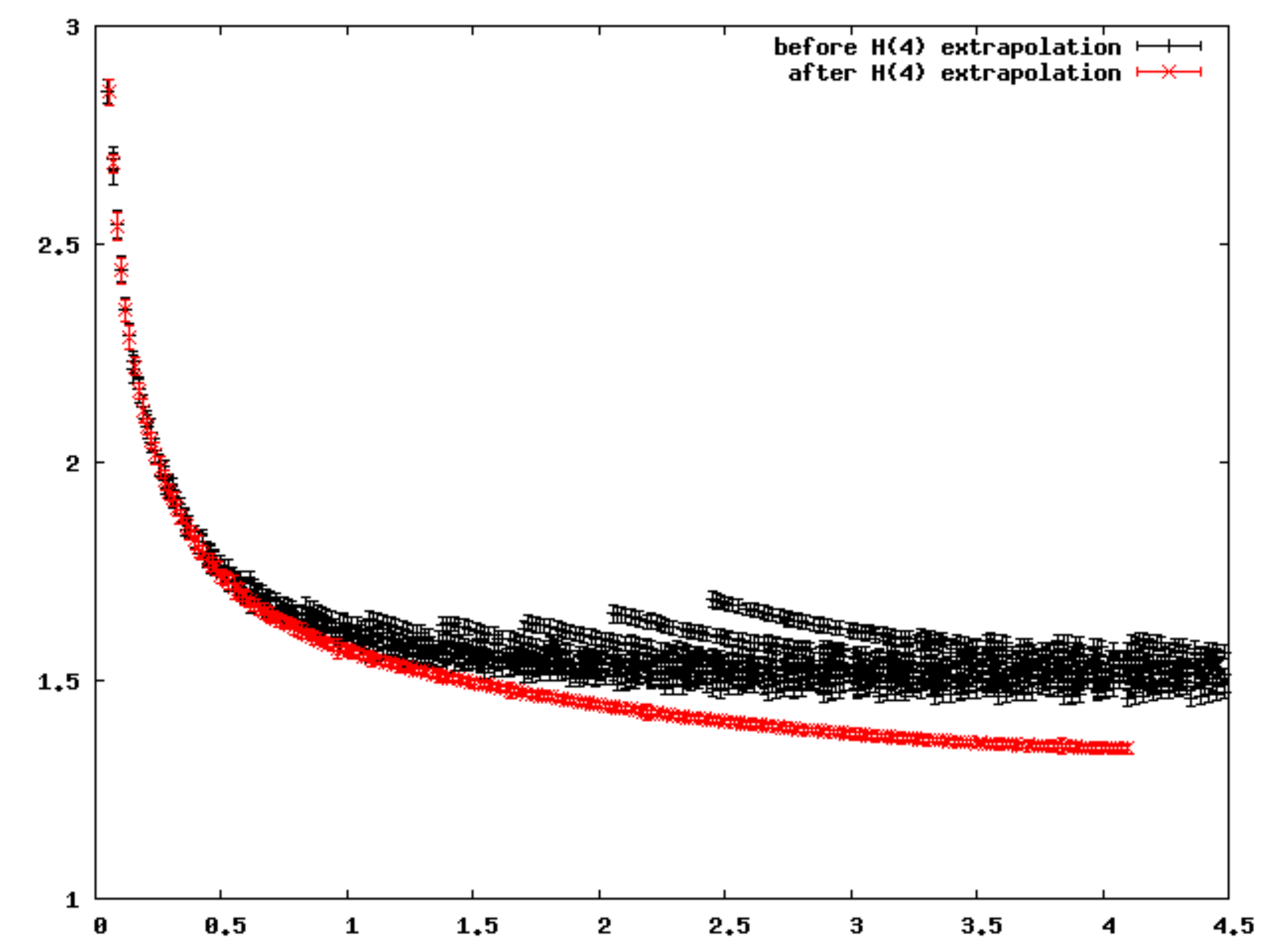}
    & 
    \includegraphics[width=9cm]{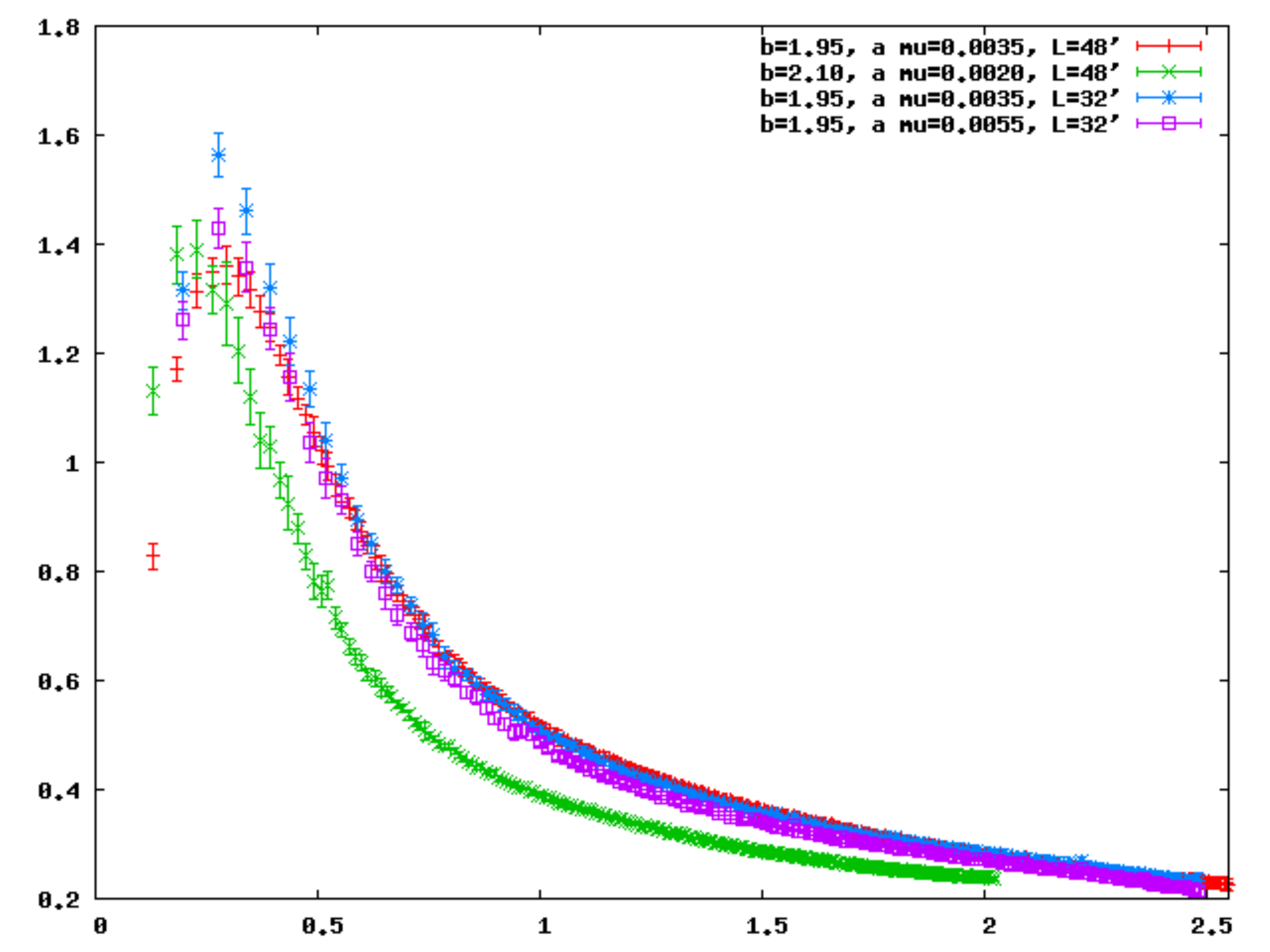}
    \\    
    $a^2(\beta) \ p^2$ &  $a(\beta) \ p$ 
    \\
    (a) & (b)
\end{tabular}
  \end{center}
\caption{\small (a) An example of the ghost dressing function lattice data (case:
$\beta=1.95$, $a \mu_l=0.035$, $L=48$)
before (black points) and after (red points) $H(4)$-extrapolation,  plotted in terms
of the square of momentum in lattice units. 
(b) The running of the coupling as a function of the momentum in lattice units for
the four ensembles of lattice data: 
$\beta=1.95$ with $a \mu_l = 0.0035$ at  $48^3\times 96$ (red) and $32^3\times 64$
lattices (blue), with 
$a\mu_l=0.0055$ at $32^3\times 64$ (violet) and $\beta=2.1$ with $a \mu_l = 0.0020$
at  $48^3\times 96$ (green).}
\label{fig:F2H4}
\end{figure}

The first stage of the analysis, as explained above, consists in the application of
the 
$H(4)$-extrapolation to cure the main type of discretization artefacts, namely the
ones 
coming from the breaking of the rotational symmetry. These effects appear to be very 
visible in Fig.~\ref{fig:F2H4}.a, where we plot the ghost dressing function before
and after $H(4)$-extrapolation 
in terms of the square of the momentum in lattice units. 
The classical "fishbone" structures generated by the different $H(4)$ orbits
corresponding 
to the same continuum momentum can be strikingly seen before applying the
extrapolation. 
Moreover, it is obvious that, had we rather applied some sort of average over a
"democratic" 
selection of the orbits at all physical momenta, the resulting ghost dressing
function would have 
shown an anomalously flat behaviour, with no indication of the perturbative
logarithm, in the 
large momentum region. All these anomalies appear to be strikingly cured by the
$H(4)$-extrapolation 
prescription. It should be noted that a very important input to apply properly such
a prescription  
comes from the many orbits at our disposal on a large momenta window (this implies
to Fourier transform 
over large momenta). It is beneficial because, for the same price, the output is a very
clean 
signal over a large momenta window that permits a very precise checking of 
the running behaviour. The results of the extrapolation for the coupling in Taylor
scheme are 
plotted against the momentum in lattice units in Fig.~\ref{fig:F2H4}.(b) for the
four lattice 
ensembles analysed.

\subsection{Flavour mass, finite volume and $O(4)$-invariant artefacts}

\begin{figure}[h]
  \begin{center}
  \begin{tabular}{cc}
    $\hat{\alpha}_T(p^2)$ \rule[0cm]{7.5cm}{0cm} & $\hat{\alpha}_T(p^2)$ \rule[0cm]{7.5cm}{0cm}
    \\
    \includegraphics[width=9cm]{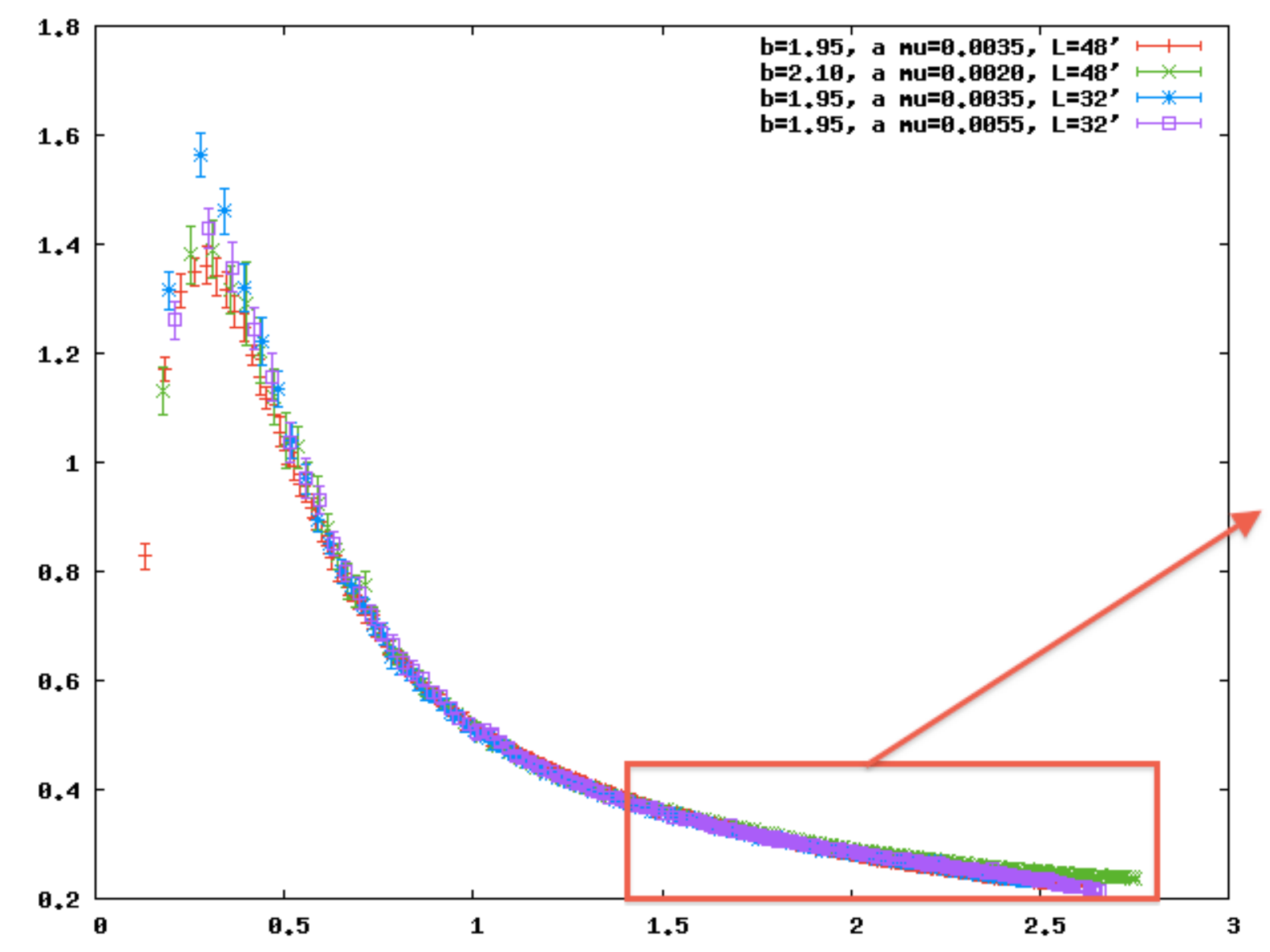}
    &
    \includegraphics[width=9cm]{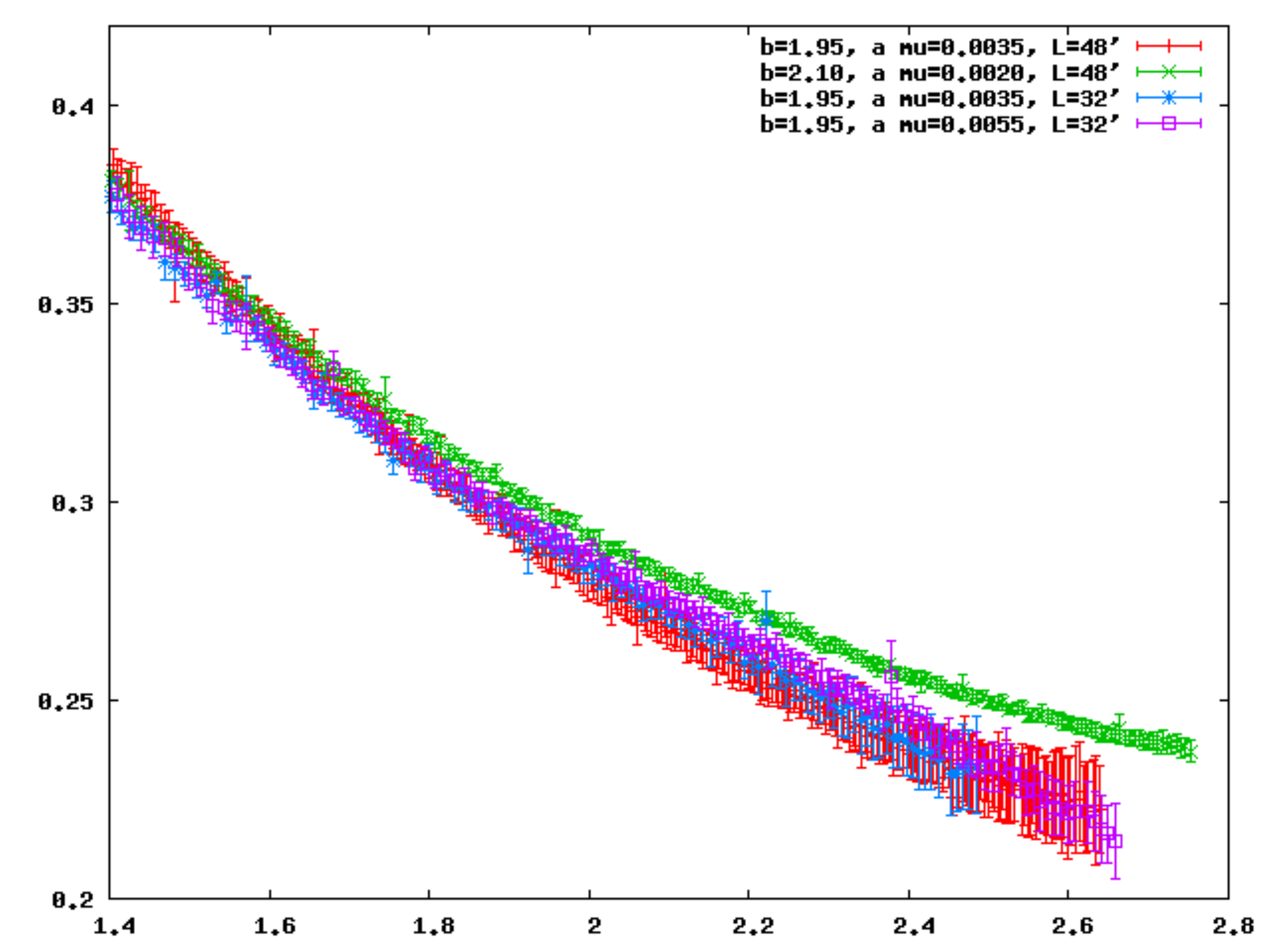}
    \\
    $a(\beta) \ p$ 
    & 
    $a(\beta)\ p$ 
\end{tabular}
  \end{center}
\caption{\small The same shown in Fig.~\ref{fig:F2H4}.(b) but after the appropriate
rescaling of the x-axis to superimpose the four 
curves as much as possible. The right plot is a zoom for the
large momentum region to  exhibit 
the discrepancies indicating that $O(4)$-invariant artefacts are present.}
\label{fig:match}
\end{figure}

After $H(4)$-extrapolation, a main part of the discretization artefacts are supposed to be removed
and, if so, all the curves for 
the running coupling in Fig.~\ref{fig:F2H4}.(b) should appear superimposed after being
re-expressed in terms of the 
momentum in physical units, at least over the momentum region where other artefacts
do not play any significant role. 
In particular, with no further conversion, the three ones for $\beta=1.95$ should
coincide with each other (and they are close to)
and the one for $\beta=2.1$ should also after the appropriate rescaling of the
momentum in coordinates axis. 
On the other hand, the comparison of the running for the three ensembles with same
$\beta$ parameter shows 
that no relevant finite volume effect happens above a lattice momentum 
of the order of $ ~ 0.5$, while a small flavour mass effect appears to be visible:
the two ensembles with same bare mass 
($a\mu_l=0.0035$) and different volumes appear nicely superimposed while the ones with
same lattice 
volume ($32^3\times 64$) and different bare masses seem to require an additional
fine rescaling. Then, we apply a rescaling 
factor to the lattice momentum, when needed, to render the four curves coincident
and show the results in Fig.~\ref{fig:match}. 
We choose to rescale all the data to those of the ensembles with $a\mu_l = 0.0035$ at $\beta = 1.95$,
since their lattice spacing seems rather safely established and we have two different volumes 
which agree fairly well. We thus 
obtain an optimal rescaling factor of $1.07$ for 
lattice momenta with $a \mu_l=0.0055$ at $\beta=1.95$ and a factor of $1.36$
(containing both the ratio of lattice 
spacings and possible flavour bare mass effects) for those with $a \mu_l=0.0020$ at
$\beta=2.1$.
The agreement is indeed impressive and make us conclude that the flavour bare
mass effect can be absorbed 
by physical calibration of the lattice spacing and be either removed by chiral
extrapolation, when possible, or included in the 
calibration systematic uncertainty. Nevertheless, it is also manifestly shown by
large lattice momentum pattern of data 
(see the right plot of Fig.~\ref{fig:match}) that the $O(4)$-invariant artefacts,
which cannot be of course cured by 
the $H(4)$-extrapolation, still survive and demand some treatment for a precise
analysis of the running. 
We will proceed to remove the remaining discretization artefacts for all our lattice
data sets by applying 
\eq{eq:a2p2} with the requirement that the coefficient $c_{a2p2}$, the correction
being a lattice artefact, 
should be universal. This will be explained in the next subsection.

\subsection{The physical running, the gluon condensate and $\Lambda_{\msbar}$}

\begin{figure}[h]
  \begin{center}
  \begin{tabular}{cc}
    \includegraphics[width=9cm]{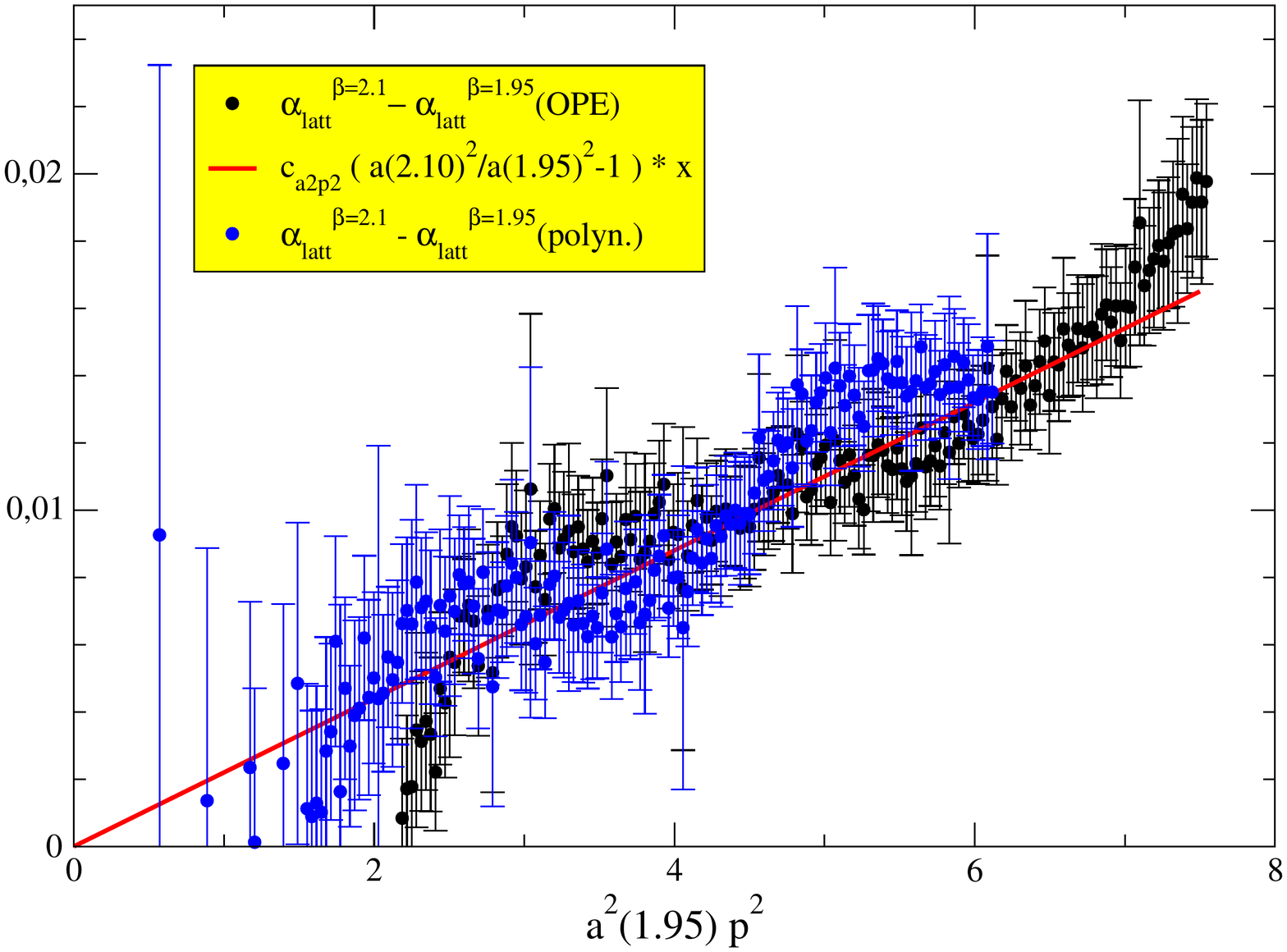}
    &
    \includegraphics[width=9cm]{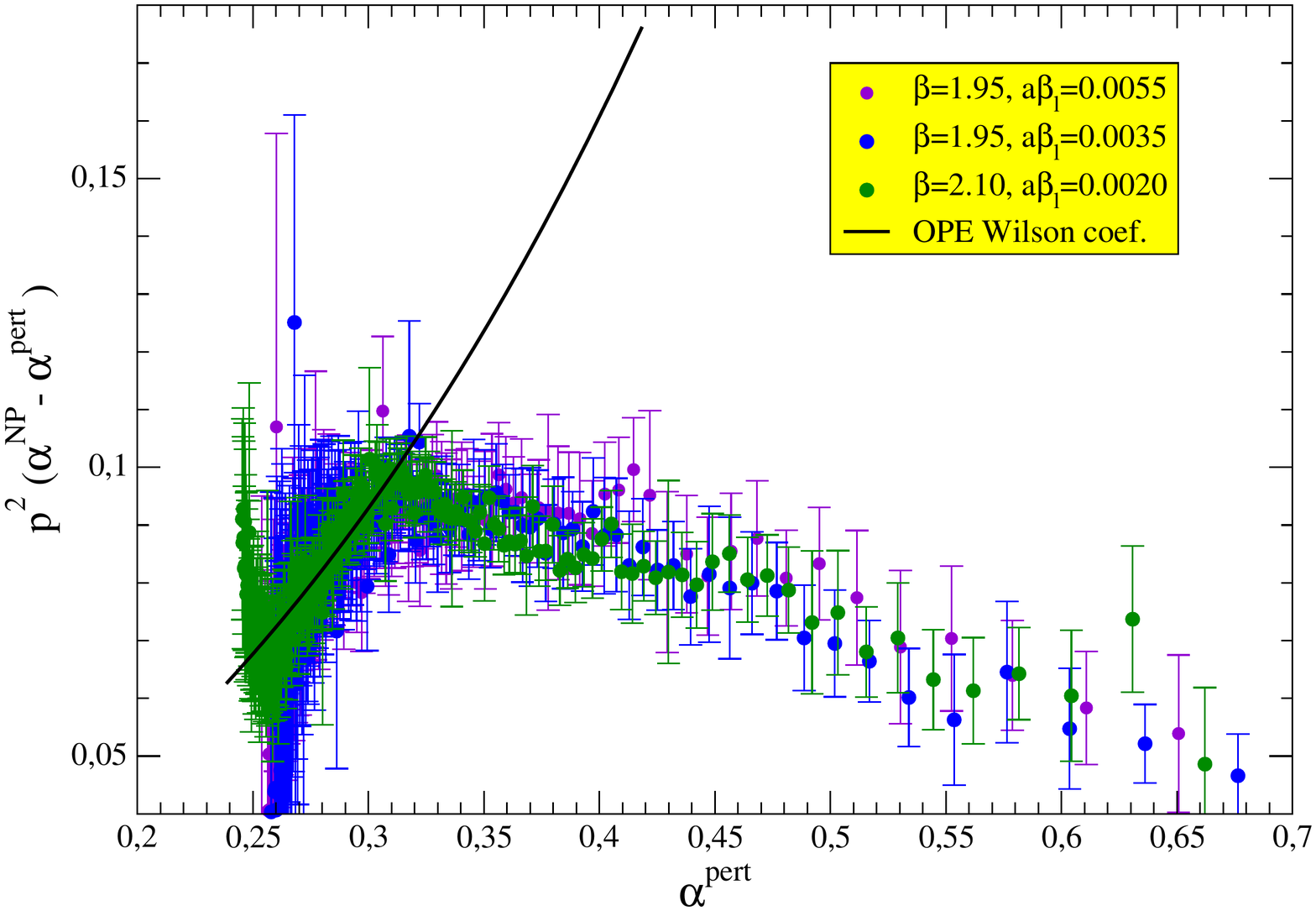} \\
    \includegraphics[width=9cm]{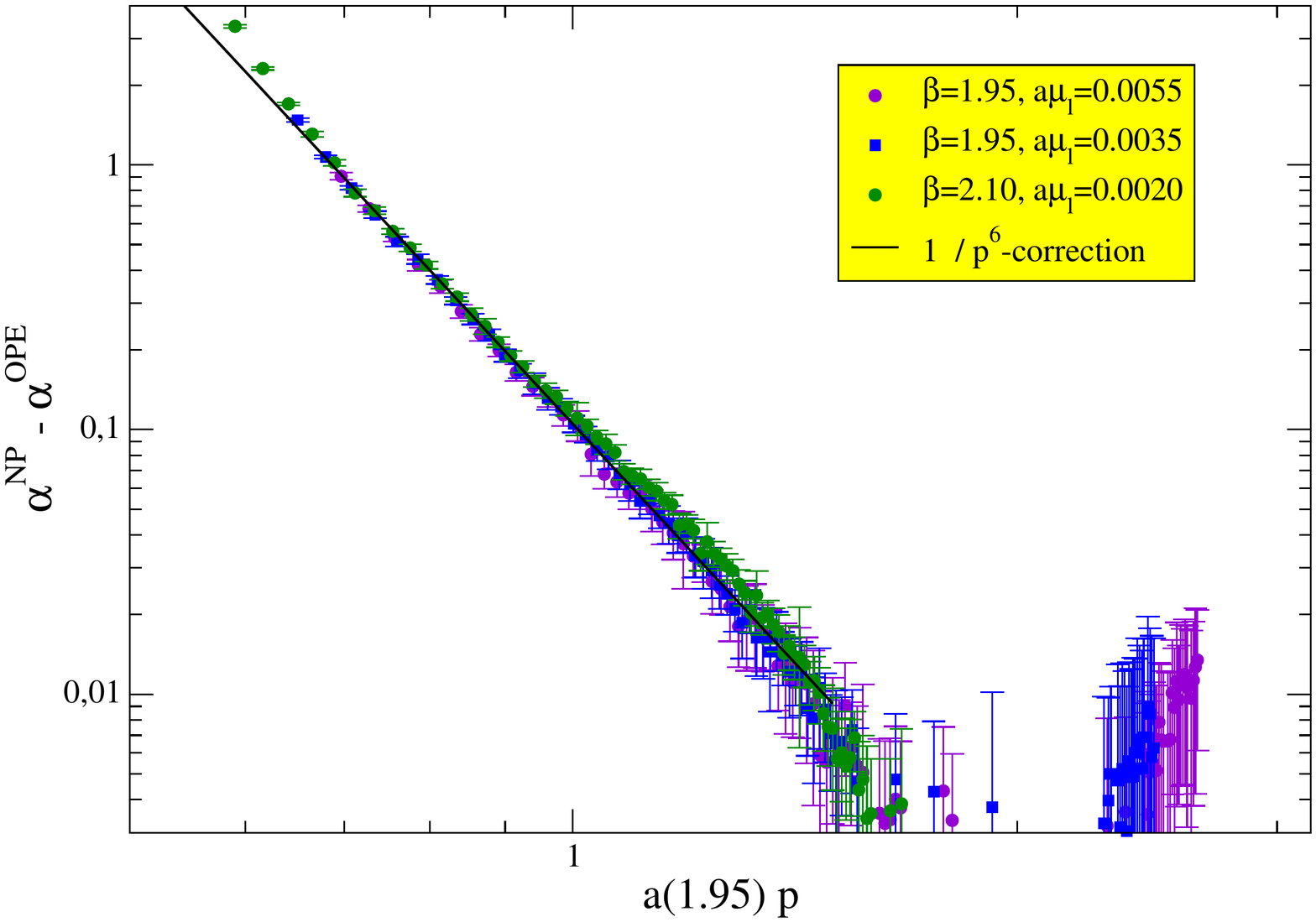}
    &
    \includegraphics[width=9cm]{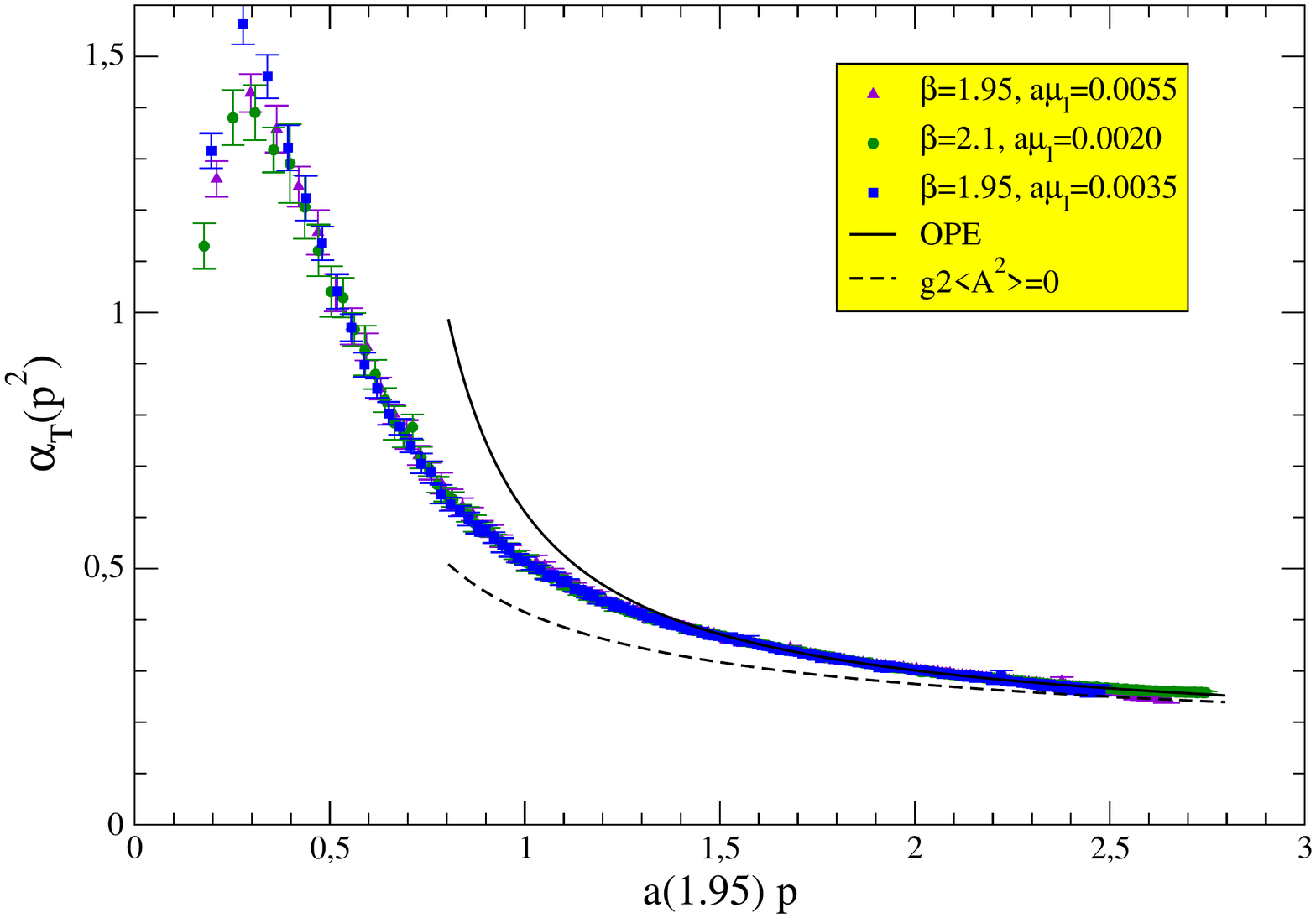}
\end{tabular}
  \end{center}
\caption{\small (Upper-left) Check of \eq{eq:checkca2p2} from the lattice data at $\beta=1.95$ ($a\mu_l=0.0035$) and $\beta=2.1$ 
($a\mu_l=0.0020$), as explained in the text. (Upper-right) deviation from the lattice data free of discretization artefacts 
with respect to the prediction of the four-loop perturbative theory, with a $\Lambda_{\msbar}$ taken from (\ref{eq:global}), 
plotted in terms of the perturbative running; the solid line shows the leading non-perturbative OPE prediction, 
\eq{eq:WilCoef}'s r.h.s. 
(Bottom-left) The departure of lattice data from the leading non-perturbative OPE prediction for the running coupling  
plotted in logarithmic scales, in terms of the momentum in lattice units of $a^{-1}(1.95)$: a next-to-leading 
$1/p^6$ behaviour is strikingly manifest. (Bottom-right) The physical running of the strong coupling obtained from the lattice 
data free of discretization artefacts, expressed in terms of the momentum in units of $a^{-1}(1.95)$; the solid line stands here 
for the best-fit with Eqs.~(\ref{eq:a2p2},\ref{alphahNP}), while the dotted one is for the four-loop perturbative prediction.}
\label{fig:plotfin}
\end{figure}

As explained in Sec.~\ref{sec:proc}, Eqs.~(\ref{eq:a2p2},\ref{alphahNP}) can be
directly applied to fit 
the lattice data plotted in Fig.~\ref{fig:F2H4}.(b) with only three free parameters.
In lattice units  they are on one hand $\Lambda_{\msbar} a(\beta)$ and
$g^2 \langle A^2 \rangle a^2(\beta)$, which depend on the lattice spacing at each
simulation and on the physical values for $\Lambda_{\msbar}$ and for the Landau-gauge gluon condensate, 
and on the other hand $c_{a2p2}$ which should be the same number for any lattice data set.
In the following, we will only analyze 
the two ensembles at $\beta=1.95$ simulated for $32^3\times 64$ lattices and 
the one at $\beta=2.1$ for a $48^4\times 96$ lattice, all of them sharing 
approximately the same lattice volume in physical units. 
The ensemble at $\beta=1.95$ for a $48^4\times 96$ lattice, for which 
we exploited a smaller number of gauge configurations and have larger statistical 
errors, has been used to check finite-size effects. 
Then, as we showed that no visible finite-size effect survives 
above $a(\beta)p \simeq 0.5$ 
and that the flavour bare mass effects can be fairly well described 
by a lattice calibration, we fit independently any ensemble of data at both
$\beta=2.1$ 
and $\beta=1.95$, by imposing the coefficient $c_{a2p2}$ to be universal, and
including a (fitted)
rescaling factor to bring the heavier mass data to superimpose with the
lighter ones at $\beta=1.95$. 
Thus, we obtain the results of Tab.~\ref{tab:res} for the best-fit
parameters. Using these values with the appropriate rescaling, 
we plot in Fig.~\ref{fig:plotfin}  the fitted running coupling obtained for the different
lattices after removing all the discretization artefacts.

\begin{table}[ht]
\begin{tabular}{|c|c|c|c|c|c|}
\hline
$\beta$ & $ a(\beta) \mu_l $ & $ a(\beta) \Lambda_{\msbar}$ & $ a(\beta)^2 g^2 \langle A^2 \rangle $ & $ a(\beta)/a(2.1) $ 
& $c_{a2p2}$  \\
\hline
1.95 & 0.0035 & 0.126(12) & 0.7(3)   & 1.36(16) & -0.0047(12)\\
     & 0.0055 & 0.117(8)  & 0.57(22) & 1.27(10) & \\
\cline{1-5}
2.1 &  0.0020 & 0.092(5) & 0.40(9) & 1 & \\  
\hline
\end{tabular}
\caption{Best-fit parameters for the confrontation of the different lattice data and Eqs.~(\ref{eq:a2p2},\ref{alphahNP}), as 
explained in the text. The local operator $A^2$ has been renormalized at $q_0 a(2.1) \simeq 3$ 
which corresponds to $q_0 \simeq 10$ GeV when the appropriate conversion is applied. 
The errors have been computed by using a jackknife procedure.}
\label{tab:res}
\end{table}

The universality of the coefficient $c_{a2p2}$ and the nature of the remaining 
discretization artefacts that could be seen in Fig.~\ref{fig:match} can be directly checked
on the data. Actually we have 
\beq\label{eq:checkca2p2}
\alpha_{\rm Latt}^{\beta=2.1}\left( a(1.95) p \right) \ - \
 \alpha_{\rm Latt}^{\beta=1.95}\left( a(1.95) p \right) \ = \ 
 \left( \frac{a^2(2.1)}{a^2(1.95)} - 1\right) \ c_{a2p2} \ a^2(1.95) p^2 \ + \ o\left( a^2(1.95) p^2 \right) \ ,
\eeq
which means that, after an appropriate rescaling implying for all the momenta to be written in units 
of the lattice spacing at a given reference $\beta$ ($\beta$ = $1.95$ and $a \mu_l=0.0035$ has been choosen here),
 the difference between the data for the coupling 
obtained at $\beta=2.1$ and those at $\beta=1.95$ has to be proportional to $c_{a2p2}$ times the square of the 
momentum when the momentum is not too large. 
This is illustrated in the upper-left plot in Fig.~\ref{fig:plotfin}, where the linear behaviour 
on $a^2p^2$ is verified (for the two lighter masses). The slope given by the r.h.s. of \eq{eq:checkca2p2} 
with the best-fit parameters of Tab.~\ref{tab:res} perfectly agrees with the data. In order to perform 
the subtraction in the l.h.s of \eq{eq:checkca2p2} some interpolation procedure is required 
to estimate the values for the momenta at $\beta=2.1$ from the data at $\beta=1.95$. To this purpose, 
we used the expression given in  Eqs.~(\ref{eq:a2p2},\ref{alphahNP})  with the  best fit parameters 
to analytically represent the data at $\beta=1.95$. The result for the 
subtraction is shown by the black points  in the plot. This procedure does not work below $a(1.95) p \simeq 1.5$, 
as the  expressions  Eqs.~(\ref{eq:a2p2},\ref{alphahNP}) do not represent properly the data in this region.
Any other interpolating formula, as far as it fits well, would provide us with 
equivalent results ; we illustrate this point in the plot by using also  a polynomial of fourth degree to describe the lattice 
data at $\beta=1.95$. The result for the subtraction in this case is given by the blue points. 
The only quantity we need to estimate the coefficient $c_{a2p2}$ from the data is the ratio of lattice spacings. 
The large uncertainty given for these quantities in Tab.~\ref{tab:res} is partially a consequence 
of our present analysis, which  uses only  the momentum window where the OPE prediction 
given by \eq{alphahNP} appears to be in order. 
As far as any UV cut-off contribution should vanish at the infinite cut-off limit for 
the running coupling defined by \eq{alpha}, all the data, properly corrected for lattice artefacts, 
from different lattice simulations should scale when expressed in terms of physical units. 
Had we only then be interested in obtaining the ratio of lattice spacings, 
the matching would be performed over a much larger momenta window\footnote{One can perfom the matching all over the region 
where finite-size effects appear not to be visible, although the matching procedure would imply some sort of 
practical fit of the lattice data inside such a region. This will be the object of a forthcoming work.} 
and the uncertainties would be drastically reduced. Anyhow, in the analysis of the present paper, we will follow 
a different fitting strategy, as will be seen in the next subsection.

\subsection{The global fit}

As we concluded in the previous subsection, the analysis of the three different lattice data sets clearly indicates that 
the flavour bare mass effects can be fairly well described by a lattice calibration. This means that we can suppose 
that the lattice spacing for any bare coupling and flavour mass can be written as
\beq
a(\beta,\mu_l) \ = \ a(\beta,0) \ \left( 1 + \ c_{a\mu_l} \ a(\beta,0)^2 \mu_l^2 + o\left( a^2\mu_l^2 \right) \right) \ ,
\eeq
where $c_{a\mu}$ gives the slope for the chiral behaviour of the lattice spacing. Of course, this light-quark bare mass dependence
for the lattice spacing must be transferred to any physical quantity like the Taylor coupling, after its lattice artefacts have 
been removed,
\beq\label{eq:chialphaT}
\alpha_T\left(\rule[0cm]{0cm}{0.35cm} a^2(\beta,\mu_l) p^2 \right) &=& \alpha_T\left(\rule[0cm]{0cm}{0.35cm} a^2(\beta, 0) p^2 \right) \ 
+ \  2 a^2(\beta,0) p^2 \ c_{a\mu} \left. \frac{d\alpha_T(x)}{dx} \right|_{x=a^2(\beta,0) p^2} \ a^2(\beta,0) \mu_l^2 \ + \ \cdots 
\nonumber \\
&\simeq& \alpha_T\left(\rule[0cm]{0cm}{0.35cm} a^2(\beta, 0) p^2 \right) \ + \ R_0\left(\rule[0cm]{0cm}{0.35cm} a^2(\beta, 0) p^2 \right) 
\ a^2(\beta,0) \mu_l^2 \ ,
\eeq
where we also assumed the strong coupling not to ``feel'' any additional light-quark bare mass effect, as it is 
clearly suggested by the results of the previous subsection. In the analysis of lattice configurations for ${\rm N}_f=2$ twisted-mass 
flavours in ref.~\cite{Blossier:2010ky}, \eq{eq:chialphaT} was successfully applied to extrapolate down to zero light-quark mass all the 
data within a narrow momentum window where $R_0$ in \eq{eq:chialphaT} was shown to be well approximated 
by a constant~\footnote{This was found to happen for $ 1 \leq ap \leq 1.5$, 
where the decreasing of the derivative of $\alpha_T$ compensated the increasing due to the factor $a^2p^2$. 
It should be furthermore noticed that  the derivative is 
negative, while $c_{a\mu}$ appears to be positive from the chiral extrapolation of the Sommer parameter, $r_0/a$, in 
ref.~\cite{Baron:2010bv}. This agrees with the sign of the chiral slope for the Taylor coupling in ref.~\cite{Blossier:2010ky}.}.
Here, we will procceed otherwise: we will take the ratios of lattice spacings, $a(\beta,a\mu_l)/a(1.95,0.0035)$, from the 
previous subsection analysis and express the Taylor coupling from the three ensembles 
of lattice data in terms of the momentum in units of $a(1.95,0.0035)^{-1}$. Then, we make a 
global fit for the three ensembles (for all momenta above $a(1.95,0.0035) \; p$=1.5) 
and obtain the following results:
\beq\label{eq:global}
\Lambda_{\msbar} \ a(1.95,0.0035) \ = \ 0.125(5) \ , \ \ a^2(1.95,0.0035) \ g^2(q_0^2) \langle A^2 \rangle_{R, q_0^2} \ = \ 0.70(6) \ , \ \ c_{a2p2} \ = \ -0.0046(7) \ ,
\eeq
whith a best $\chi^2=103.8$ for 317 degrees of freedom. Some plots resulting from the global fit 
can be seen in Fig.~\ref{fig:plotfin}. 

The upper-right plot of Fig.~\ref{fig:plotfin} shows the lattice data, after the subtraction of the 
perturbative running and the $O(4)$-invariant artefacts, multiplied then by the square of the momentum in units of $a^{-1}(1.95)$  
and plotted in terms of the four-loop perturbative value of the coupling at the same momentum. 
To obtain the perturbative coupling, we apply the value of $\Lambda_{\msbar}$ obtained from (\ref{eq:global}) which, 
after being converted to physical units with lattice spacing taken from ref.~\cite{Baron:2010bv,Baron:2011sf}, appears to be in very good 
agreement with the experimental result (see below). According to \eq{alphahNP}, one would obtain: 
\beq\label{eq:WilCoef}
p^2 \ \left(\alpha_T(p^2) - \alpha^{\rm pert}_T(p^2)\right)
\ = \
\
\frac{9 g^2_T(q_0^2) \langle A^2 \rangle_{R,q_0^2}} {4 (N_C^2-1)}
\ R\left(\alpha^{\rm pert}_T(p^2),\alpha^{\rm pert}_T(q_0^2) \right) 
\ \alpha^{\rm pert}_T(q_0^2) \left( \frac{\alpha^{\rm pert}_T(p^2)}{\alpha^{\rm pert}_T(q_0^2)}
\right)^{2-\gamma_0^{A^2}/\beta_0} \ .
\eeq
The solid line in the upper-right plot of Fig.~\ref{fig:plotfin} corresponds to the r.h.s of \eq{eq:WilCoef} with 
the value for the Landau-gauge gluon condensate taken from \eq{eq:global}. 
One should notice that the departure from zero for the lattice data in the plot can be only explained by 
non-perturbative contributions. Furthermore, the Wilson coefficient for the Landau-gauge gluon condensate 
in the OPE expansion successfully accounts for the non-flat behaviour from the lattice data 
in the small coupling regime. This provides with a striking indication that (and where) the OPE analysis 
is in order. 
Next, in the bottom-left plot of Fig.~\ref{fig:plotfin}, we show the departure of the lattice data 
from the prediction given by Eqs.~(\ref{eq:a2p2},\ref{alphahNP}), plotted in terms of the momentum 
in units of $a^{-1}(1.95)$, with logarithmic scales for both axes. The data seem to indicate that the next-to-leading 
non-perturbative correction is highly dominated by an  $1/p^6$ term. This is just a factual statement which might suggest 
either that the $1/p^4$ OPE contributions are negligible when compared with the $1/p^6$ ones or that the product of 
the leading $1/p^4$ terms and the involved Wilson coefficients leave with an effective $1/p^6$ behaviour. One might also 
guess that a different non-perturbative mechanism dominates over the momentum window where the dimension-four 
OPE condensates had to be visible. 
Finally, the bottom-right plot shows the physical running of the coupling and how well the OPE formula with 
the results from \eq{eq:global} fits the data.

\subsection{Conversion to physical units}

The two main purposes of this paper are to show the impact of the OPE power corrections in describing the running of 
the strong coupling and to give an estimate for $\Lambda_{\msbar}$ from lattice QCD simulations with a dynamical charm quark. 
The results for both goals can be summarized by the conversion to physical units for $\Lambda_{\msbar}$ and the Landau-gauge 
gluon condensate. To this purpose, we will apply
\beq\label{eq:abs}
a(1.95,0.0035) \ = \ \frac{a(1.95,0.0035)}{a(1.95,0)} \ a(1.95,0) 
\eeq
where the absolute calibration of the lattice spacing at $\beta=1.95$, after 
the chiral extrapolation for the light quark mass, will be taken
from refs.~\cite{Baron:2010bv,Baron:2011sf}: $a(1.95,0)=0.0779(2)$ fm; and where 
we approximate
\beq\label{eq:applat}
\frac{a(1.95,0.0035)}{a(1.95,0)} \ = \ 1^{+0.03}_{-0} \ .
\eeq
The systematic error quoted here has been estimated from the 
chiral extrapolation of the Sommer parameter in ref.~\cite{Baron:2010bv}. There, as can be seen in 
plot 6.(b), one gets
\beq\label{eq:r0}
\frac{(r_0/a)^{a\mu_l=0}}{(r_0/a)^{a\mu_l=0.0035}} \ - \ 1 \ = \  0.015(11) \ .
\eeq
Then, if the string tension for the static interquark potential is supposed not to depend 
very much on the light quark mass, \eq{eq:r0} gives the conservative systematic uncertainty for 
the deviation from 1 in \eq{eq:applat}. Thus, we apply Eqs.~(\ref{eq:abs}-\ref{eq:r0}) into 
\eq{eq:global} and obtain:
\beq
\Lambda_{\msbar}^ {N_f=4} &=&  316 \pm13\pm 8 ^{+0}_{-9} \ \mbox{\rm MeV}  \nonumber \\
g^2(q_0^2) \langle A^2 \rangle_{R, q^2_0} &=& 4.5 \pm 0.4 \pm 0.23 ^{+0}_{-0.3} \ \mbox{GeV}^2 \ ;
\eeq
where the first quoted error is statistical, the second one reflects the present 
uncertainty on the absolute calibration of the lattice spacing that 
we roughly (and conservatively) estimate to be of $\pm 2.5$ \%; and the 
third one is for the chiral extrapolation of the light quark mass. 
More precise estimates for these systematics uncertainties will be accessible 
with more data (more simulations at different $\beta$'s and for more light quark masses).
Finally,  the value for $\Lambda_{\msbar}^{N_f=4}$ and the four-loop perturbative running with the appropriate crossing  
of the bottom mass threshold at $m_b(m_b)=4.19^{+18}_{-6}$GeV~\cite{Nakamura:2010zzi} 
can be used to estimate the value of the coupling at the $Z^0$ mass, 
\beq
\alpha_S\left(M_{Z^0}\right) \ = \ 0.1198(9)(5)^{+0}_{-5} \ ;
\eeq
where the errors have been properly propagated. 
This is a first result that will be refined, mainly by improving the precision 
for the estimates of systematic uncertainties. However, it appears to be pretty compatible with the last 
world average given by PDG~\cite{Nakamura:2010zzi}: 0.1184(7). 
Although a more detailed comparison of our result with PDG average 
and a discussion of its implications will be left for a 
phenomenologically targeted forthcoming letter, we should remark that our result including 
strange and charm dynamical quarks ($N_F=$2+1+1) appears to be slightly larger than 
the lattice estimate for ${\rm N}_f$=2+1 staggered fermions, applied to obtain the PDG 
average: $\alpha_S\left(M_{Z^0}\right)=0.1183(8)$~\cite{Davies:2008sw}. 
Assuming no systematic effect to appear from the different fermion actions, 
the meaning for the 1-$\sigma$ discrepancy of both central values, if any,  
can be explained from the procedure applied to cross the 
threshold from ${\rm N}_f$=2+1 to ${\rm N}_f$=2+1+1 flavours. 
That procedure is very well established and controlled in perturbation 
theory~\cite{Nakamura:2010zzi,Chetyrkin:2005ia}, 
but some non-perturbative effects may still appear at the charm quark running mass. 

Indeed, if we compare the result of $\Lambda_{\overline{\rm MS}}$ for ${\rm N}_f=2$ in ref.~\cite{Blossier:2010ky}, 
the central value ranging from 310 to 330 MeV (depending on the lattice size calibration 
at different values of $\beta$), with that for ${\rm N}_f$=2+1+1 in this paper, it can be 
concluded that the effect for the running coupling of crossing the strange or charm 
quark thresholds is not very significant\footnote{In Sec. 4.5 of 
ref.~\cite{Blossier:2010ky}, a significant difference between  quenched (${\rm N}_f=0$) 
and ${\rm N}_f=2$ lattice results for  $\Lambda_{\overline{\rm MS}}$ (once lattice 
spacing is calibrated from $f_{\pi}$) was observed. In that case, simulations for 
infinite (quenching) and vanishing (chiral limit) quark flavours were compared, 
although it is well known that the chiral lmit for the quenched case is wrong. 
On the contrary, when comparing ${\rm N}_f$=2+1+1 and ${\rm N}_f$=2 
(or ${\rm N}_f$=2+1) cases, we deal with infinitely massive flavours (s and c) in the latter 
and heavy (c) or mid-heavy (s) in the former. Then, not to see the same significant effect 
cannot be too surprising.} on $\Lambda_{\overline{\rm MS}}$. 
On the other hand, applying just the perturbative recipe to cross the charm quark 
threshold would result in a stronger decreasing of $\Lambda_{\overline{\rm MS}}$ from 
${\rm N}_f$=2+1 to ${\rm N}_f$=2+1+1 flavours. Thus, an enhancement for the 
estimate of $\alpha_S\left(M_{Z^0}\right)$ obtained from the 
${\rm N}_f$=2+1+1 lattice result for $\Lambda_{\overline{\rm MS}}$  can be understood when 
comparing to the one obtained by applying that perturbative recipe to cross the charm quark 
threshold with the ${\rm N}_f$=2+1 lattice result.


\section{Conclusions}

We used lattice gauge field configurations with two degenerate light and one heavy doublet 
of twisted mass flavors, produced within the framework of ETM collaboration, to compute the 
running strong coupling in the MOM Taylor scheme. In this particular renormalization scheme, 
the lattice computation of this coupling has the very nice feature of involving only propagators. 
This allows for a very precise control of the lattice artefacts and other systematic uncertainties. 
In particular, the so-called $H(4)$-extrapolation procedure which has already been proved 
to be very effective at eliminating the discretization artefacts of two-point correlation functions
is efficiently at work in this analysis. The dominant $O(4)$ artefact is also rather easily isolated 
and eliminated.
On the other hand, the renormalized coupling, defined in MOM taylor scheme by 
the combination of the ghost and gluon bare propagators, must only depend on 
the UV cut-off through residual contributions vanishing at the infinite cut-off limit.  
Thus, the Taylor coupling computed from the lattice   
must join the continuum prediction at this infinite cut-off limit. 
This is also a strong point to cure properly lattice artefacts and get reliable results. 
Thus, we evaluated the running strong coupling  over a rather 
large window of lattice momenta and, after the appropriate relative calibration, 
confront it with the perturbative prediction available at the four-loop level. 
We clearly demonstrate the necessity to include non-perturbative power corrections to
get an accurate description for the  behaviour of the coupling constant. We
then show these corrections follow the OPE predictions
when a non-vanishing dimension-two Landau-gauge gluon condensate is present. This is to our knowledge 
the first time a Wilson coefficient is directly confronted to numerical results. 
Higher order terms are visible and seem to be dominated by $1/p^6$ contributions instead 
of the expected $1/p^4$ ones.

The precise comparison of lattice estimates with continuum formula allows the estimate of 
both the gluon condensate and $\Lambda_{\msbar}$. The latter is in a very good agreement 
with the world average of its experimental determinations provided by PDG, 
confirming that the picture we advanced, first by the analysis of quenched lattice data and 
next by studying lattice simulations with two twisted-mass flavours, agrees very well with the 
real world when a full QCD analysis is performed including a heavy doublet for strange and charm 
quarks. Going the other way around,  with  the experimental value for $\Lambda_{\msbar}$
as an input, this approach could be used to provide quite  good absolute
determinations of  the lattice spacings.
Relative measurements (ratio of lattice spacings)
through the superimposition of the different curves for $\alpha(a^2p^2)$ over a large window 
can be also obtained with a good level of precision. 
The perturbative running from our results up to the Z boson mass is well known
in perturbation theory and we apply the standard formula~\cite{Nakamura:2010zzi} to cross the b quark threshold.
{\it We do not need to consider the charm quark threshold since, for the first time to our knowledge,
we have used dynamical charm in our computation. This is a very significant gain : the crossing
of the charm threshold using perturbative QCD is quite questionable, since, as we have seen,
non-perturbative effects (OPE power terms) are sizeable at this energy.} 

Further works are in progress. In particular data with different masses at ${\rm N}_f$=2+1+1
and from simulations with ${\rm N}_f=4$ light flavours will soon be  available 
to help to improve on the question of the mass effect.

\section*{Acknowledgements} 
We are particularly indebted to A. Le Yaouanc, J. P. Leroy 
and J. Micheli for participating in many fruitful discussions at 
the preliminar stages of this work and to S.~F.~Reker and G.~C.~Rossi for their very 
attentive reading of the manuscript and valuable comments. 
We also thank the IN2P3 Computing Center 
(CNRS-Lyon), the apeNEXT computing laboratory (Rome)
and the computing center at IDRIS (CNRS-Orsay) for provinding  numerical resources. 
J. R-Q is indebted to the Spanish MICINN for the 
support by the research project FPA2009-10773 and 
to ``Junta de Andalucia'' by P07FQM02962.



\end{document}